\documentclass[preprint,onecolumn,nofootinbib]{revtex4}
\pdfoutput=1
\usepackage[colorlinks=true,linkcolor=blue,urlcolor=blue,filecolor=black,citecolor=red,pdfstartview=FitV,pdftitle={},pdfsubject={},pdfkeywords={},pdfpagemode=None,bookmarksopen=true]{hyperref}
\usepackage{graphicx}
\usepackage{amsmath}
\usepackage{amsfonts}
\usepackage{amssymb}
\usepackage{color}%
\usepackage{dcolumn}
\newcommand{\f}{\begin{equation}}
\newcommand{\ff}{\end{equation}}
\newcommand{\fa}{\begin{eqnarray}}
\newcommand{\ffa}{\end{eqnarray}}

\begin{document}
\title{Doped holographic fermionic system}
\author{Wenjun Huang $^{1}$}
\author{Guoyang Fu $^{2}$}
\author{Dan Zhang $^{1}$}
\author{Zhenhua Zhou $^{3}$}
\author{Jian-Pin Wu $^{2}$}
\thanks{Corresponding author, jianpinwu@yzu.edu.cn}
\affiliation{
$^1$ Department of Physics, School of Mathematics and Physics, Bohai University, Jinzhou 121013, China\ \\
$^2$ Center for Gravitation and Cosmology, College of Physical Science and Technology, Yangzhou University, Yangzhou 225009, China\ \\
$^3$ School of Physics and Electronic Information, Yunnan Normal University, Kunming, 650500, China
}
\begin{abstract}

  We construct a two-current model. It includes two gauge fields, which introduce the doping effect, and a neutral scalar field.
  And then we numerically construct an AdS black brane geometry with scalar hair.
  Over this background, we study the fermionic system with the pseudoscalar Yukawa coupling.
  Some universal properties from the pseudoscalar Yukawa coupling are revealed. In particular, as the coupling increases,
  there is a transfer of the spectral weight from the low energy band to the high energy band.
  The transfer is over low energy scales but not over all energy scales.
  The peculiar properties are also explored. The study shows that with the increase of the doping,
  the gap opens more difficult. It indicates that there is a competition between the pseudoscalar Yukawa coupling and the doping.

\end{abstract}
\maketitle
\section{Introduction}

The holographic duality \cite{Maldacena:1997re,Gubser:1998bc,Witten:1998qj,Aharony:1999ti}, also referring to AdS/CFT (Anti-de-Sitter/Conformal field theory) correspondence,
relates a lower-dimensional quantum field theory (QFT) to a classical gravitational system in higher dimensions.
It provides us a way to understand strongly-coupled condensed matter systems by constructing a simple gravitational dual model.
Also, it can provide some physical insight into the associated mechanisms of these strongly-coupled systems as well as the universality class of them.
The well-known examples include the holographic superconductor \cite{Hartnoll:2008vx}, holographic metal insulator phase transition (MIT) \cite{Donos:2012js,Ling:2014saa}
and (non-) Fermi liquid \cite{Liu:2009dm}.

Doping plays an essential role in condensed matter system. The temperature-doping phase diagram is landscape of exotic states of matter,
including antiferromagnetic (AF) phase, pseudogap phase, strangle metal phase and superconducting phase.
How to model the phase diagram and provide a preliminary understanding on it in holography has been an emergent and important topic of research.
Some pioneer works have already been devoted to this issue, for example \cite{Kiritsis:2015hoa,Baggioli:2015dwa}.
In these works, they depict the different phases by the properties of the electronic transport.

Also, we can study the fermion spectral function by adding probe fermions in a gravitational background.
A variety of unexpected emergent phenomena have been observed.
In \cite{Liu:2009dm}, they devoted to the study of the fermionic spectrum over RN-AdS background.
The study demonstrates that the fermionic spectrum over RN-AdS background exhibits quasi-particle-like excitation around Fermi level.
In particular, non-linear dispersion relation emerges, which is the peculiar characteristic of non-Fermi liquid system.
It is consistent with the overall picture of some unconventional phases, such as the normal state of the high-$T_c$ superconducting cuprates,
and metals close to a quantum critical point (QCP).
Some generalized studies have been widely explored, see \cite{Wu:2011bx,Wu:2011cy,Gursoy:2011gz,Alishahiha:2012nm,Fang:2012pw,Li:2012uua,Wang:2013tv,Wu:2013xta,Kuang:2014pna,Fang:2014jka,Fang:2015vpa} and therein.

The above works on the holographic fermionic systems only focus on the minimal coupling among the fermions, gauge fields and gravity.
In bottom-up approaches, beyond minimal coupling, we can explore the fermionic response of the holographic systems,
in which the fermions are coupled to gravity and gauge fields in a variety of ways.
It is certainly desirable to study how the fermionic spectrum is modified by such non-minimal bulk coupling in the dual boundary field theory,
or the novel interesting emergent phenomena induced by the non-minimal coupling.
Along this direction, the pioneer work has been implemented in \cite{Edalati:2010ww,Edalati:2010ge}, in which they introduce the dipole coupling
between the fermions and gauge fields. An interesting phenomenon that a dynamical energy gap emerges.
Especially, they find that the spectral weight transfers over all energy scale \cite{Edalati:2010ww,Edalati:2010ge}.
The emergence of dynamical gap and spectral weight transfer are the key properties of Mott insulating phase.
Lots of studies over more general geometries in \cite{Wu:2012fk,Kuang:2012ud,Wen:2012ur,Kuang:2012tq,Wu:2013oea,Wu:2014rqa,Kuang:2014yya,
Vanacore:2015,Fan:2013zqa,Fang:2015dia,Wu:2016hry,Kuang:2017bst,Fang:2016wqe,Fang:2013ixa,Seo:2018hrc}
are implemented, which confirm the robustness of the two key features found in RN-AdS background in \cite{Edalati:2010ww,Edalati:2010ge}.
Another novel non-minimal bulk coupling, the so-called pseudoscalar Yukawa coupling, in which the the spinor field coupled to a neutral scalar field,
is introduced in holographic system and the corresponding fermionic spectral function is explored in \cite{Wu:2019orq}.
In \cite{Wu:2019orq}, we also observe the emergence of a gap around the Fermi level,
which indicates an insulating phase \cite{Wu:2019orq}. Further studies shows that the insulating phase of the holographic system with pseudoscalar Yukawa coupling
is different from the Mott physics \cite{Wu:2019orq}.

However, we would like to point out that the emergence of the energy gap studied above is from the bulk coupling, i.e., dipole coupling or pseudoscalar Yukawa coupling,
but not related to the physics of localization/scattering over impurities/umklapp scattering.
Nonetheless, in this way, various emergent IR characteristics have been identified, for example, see \cite{Iqbal:2011in,Iqbal:2011aj,Faulkner:2011tm,Edalati:2010ge,Wu:2013xta,Wu:2013vma}.

In this paper, we intend to implement the temperature-doping phase diagram in holographic femrionic system.
To this end, we follow the idea of holographic effective theory to construct a model including
two gauge fields and a neutral scalar field. Therefore, we have two independently conserved currents,
which relates to different kinds of charge in the dual field theory.
The ratio of the two charges plays the role of doping \cite{Kiritsis:2015hoa}.
Over this background, we introduce the pseudoscalar Yukawa coupling as \cite{Wu:2019orq} and study its response.
This paper is organized as what follows.
In Section \ref{sec-frame}, we introduce the holographic framework.
Section \ref{sec-NR} presents the numerical results.
The conclusion and discussion are presented in Section \ref{sec-con}.

\section{Holographic framework}\label{sec-frame}

\subsection{Background geometry}\label{subsec-bg}

We consider the following action
\begin{equation}
S = \frac{1}{2{\kappa}^2}\int \mathrm{d}^{4}x\sqrt{-g}\left(R +\frac{6}{L^2}+ \mathcal{L}_M +\mathcal{L}_{\Phi}\right),
\label{s}
\end{equation}
where $\mathcal{L}_M$ and $\mathcal{L}_{\Phi}$ are given in what follows,
\begin{subequations}
\label{LMpHI}
\begin{align}
&
\mathcal{L}_M=-\frac{1}{4} Z_A(\Phi) F_{\mu\nu}F^{\mu\nu}-\frac{1}{4}Z_B(\Phi) G_{\mu\nu}G^{\mu\nu}-\frac{1}{2}Z_{AB}(\Phi) F_{\mu\nu}G^{\mu\nu}\,,
\label{LM}
\\
&
\mathcal{L}_{\Phi}=-\frac{1}{2}\nabla_{\mu}\Phi \nabla_{\nu}\Phi - \frac{1}{2}m^2\Phi^2\,.
\label{LPhi}
\end{align}
\end{subequations}
The coupling functions $Z_A(\Phi)$, $Z_B(\Phi)$ and $Z_{AB}(\Phi)$ are chose as the following specific forms
\begin{align}
Z_A(\Phi)=1+\frac{1}{2}\alpha \Phi^2\,,\ \ \ Z_B(\Phi)=1+\frac{1}{2}\beta \Phi^2\,,\ \ \ Z_{AB}(\Phi)=\frac{1}{2}\gamma\Phi^2\,.
\end{align}
This theory includes two gauge fields $A$ and $B$ and the corresponding field strengths are $F=dA$ and $G=dB$, respectively.
$\Phi$ is a neutral scalar field, which has the dimension $\Delta$ relating to the mass of the scalar field as $m^2L^2=\Delta(\Delta-3)$.

The equations of motion (EOMs) can be straightforward derived from the action \eqref{s} as
\begin{subequations}
\label{eom}
\begin{align}
&
\label{eom-g}
R_{\mu\nu}-\Big(\frac{1}{2}R+\frac{3}{L^2}\Big)g_{\mu\nu}
+\frac{1}{2}(T^A_{\mu\nu}+T^{\Phi}_{\mu\nu})=0\,,
\
\\
&
\nabla_{\mu}\left(Z_A(\Phi)F^{\mu\nu}+Z_{AB}(\Phi)G^{\mu\nu}\right)=0\,,
\
\\
&
\nabla_{\mu}\left(Z_B(\Phi)G^{\mu\nu}+Z_{AB}(\Phi)F^{\mu\nu}\right)=0\,,
\
\\
&
\Big(\nabla^2-m^2-\Big(\frac{1}{4}\alpha F^2 + \frac{1}{4}\beta G^2 +\frac{1}{2}\gamma F_{\mu\nu}G^{\mu\nu}\Big)\Big)\Phi=0\,,
\end{align}
\end{subequations}
where
\begin{subequations}
\label{em-tensor}
\begin{align}
&
\label{T-A}
T^A_{\mu\nu}=-\mathcal{L}_M g_{\mu\nu}-Z_{A}(\Phi)F_{\mu\rho}F_{\nu}^{\ \rho}-Z_{B}(\Phi) G_{\mu \rho}G_{\nu}^{\ \rho}-2 Z_{AB}(\Phi) F_{(\mu |\rho|}G_{\nu)}^{\ \rho}\,,
\
\\
&
T^{\Phi}_{\mu\nu}=-\mathcal{L}_{\Phi} g_{\mu\nu} - \nabla_{\mu}\Phi \nabla_{\nu}\Phi\,.
\end{align}
\end{subequations}
The symmetry bracket in $T^A_{\mu\nu}$ means $A_{(\mu\nu)}=(A_{\mu\nu}+A_{\nu\mu})/2$.

We shall numerically construct the black brane solution with non-trivial scalar profile.
We assume the following ansatz\footnote{Here and in what follows, we shall set $L=1$.}
\begin{subequations}
\label{bs}
\begin{align}
ds^2&={1\over
u^2}\Big[-(1-u)pUdt^2+\frac{du^2}{(1-u)pU}+Vdx^2+Vdy^2\Big]\,,\label{ansatz-metric} \\
A&=\mu(1-u)a dt\,,\label{ansatz-A} \\
B&=\delta\mu(1-u)b dt\,,\label{ansatz-A} \\
\Phi&=u^{3-\Delta}\phi\,,\label{ansatz-phi}
\end{align}
\end{subequations}
where $p(u)=1+u+u^2-\mu^2u^3/4$. $\mu$ and $\delta\mu$ are the chemical potentials of the dual boundary field theory of the gauge fields $A$ and $B$.
The two controllable chemical potentials or charge densities induce the unbalance of numbers \footnote{Two gauge fields are introduced to simulate
unbalanced mixtures in holography, which can be found in \cite{Erdmenger:2011hp,Bigazzi:2011ak,Musso:2013rnr,Amoretti:2013oia,Alsup:2012ap,Alsup:2012kr} and therein.
In addition, in \cite{Ling:2015exa}, two gauge fields are also introduced to study MIT.}.
The ration $\chi=\frac{\delta\mu}{\mu}$ represents the amount of charged impurities,
which is introduced to simulate the doping \cite{Kiritsis:2015hoa}.

$U$, $V$, $a$ and $b$ are the function of the radial coordinate $u$ only.
The temperature of the dual boundary field theory is
\fa
\label{tem}
\hat{T}=\frac{(12-\mu^2)U(1)}{16\pi}\,.
\ffa
And then we can define a scaling-invariant temperature $T\equiv \hat{T}/\mu$.
To support an asymptotic AdS black bran, we set the boundary conditions as $U(0)=V(0)=a(0)=b(0)=1$ at the UV boundary.
It is easy to derive the asymptotic behavior of $\phi(u)$ at UV boundary, which follows
\fa
\label{asy-beh-phi}
\phi(u)=\phi_0+\phi_1 u^{2\Delta-3}\,.
\label{phiu-asy}
\ffa
We identify $\phi_0$ and $\phi_1$ as the source and expectation, respectively.
The source $\phi_0$ corresponds to the coupling of the boundary QFT and deforms it.

For given scalar field mass $m^2$ and the coupling parameters $\alpha$, $\beta$ and $\gamma$, the system is determined by three scaling-invariant parameters,
namely, the Hawking temperature $T$, the doping $\chi$ and the coupling $\lambda\equiv\phi_0/\mu^{3-\Delta}$.
We mainly focus on the effect of the doping, so we turn off
$\phi_0$, i.e., setting $\phi_0=0$. In addition, we fix the parameters as $\alpha=5$, $\beta=-1$, $\gamma=1$ and $\Delta=2$ through this paper.

\begin{figure}
\center{
\includegraphics[scale=0.6]{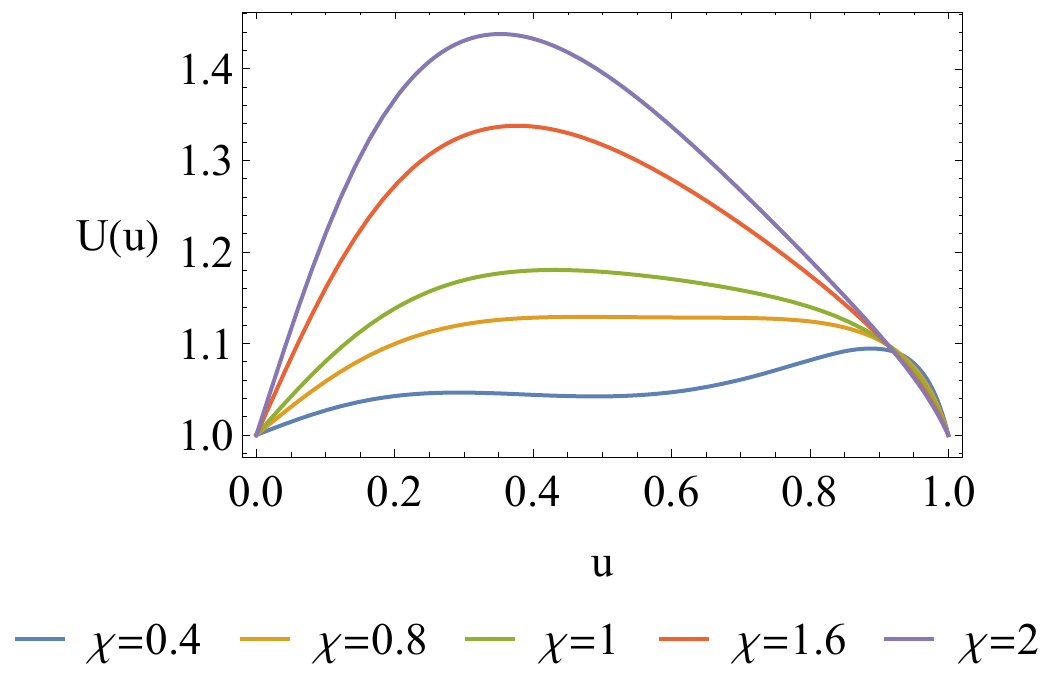}\ \hspace{0.8cm}
\includegraphics[scale=0.6]{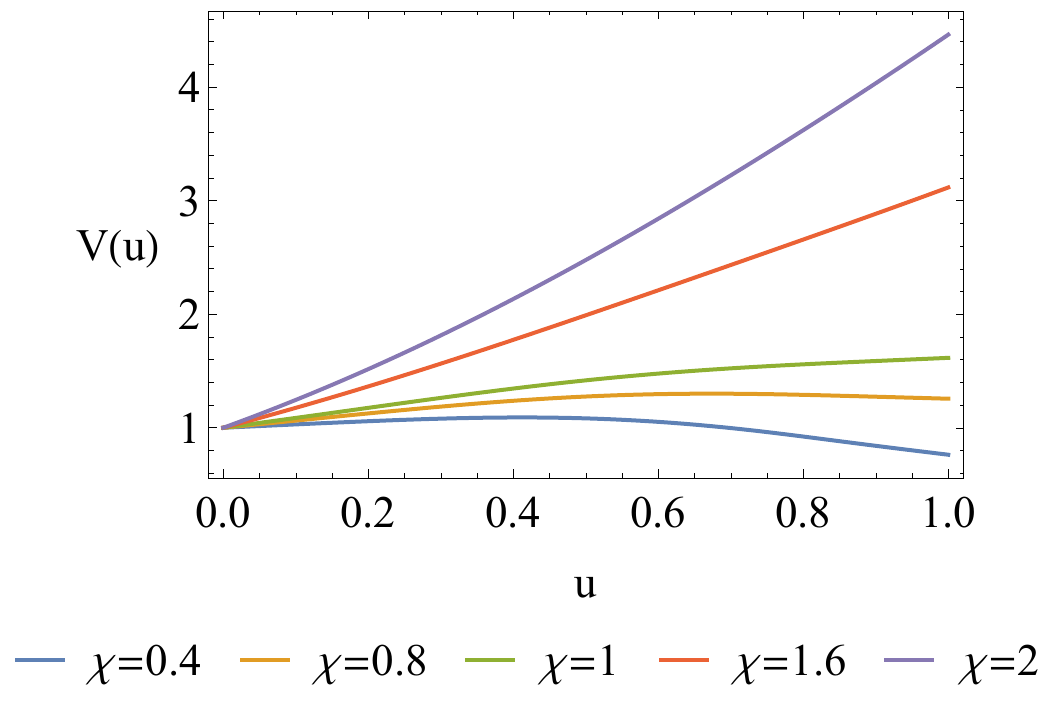}\ \\
\includegraphics[scale=0.6]{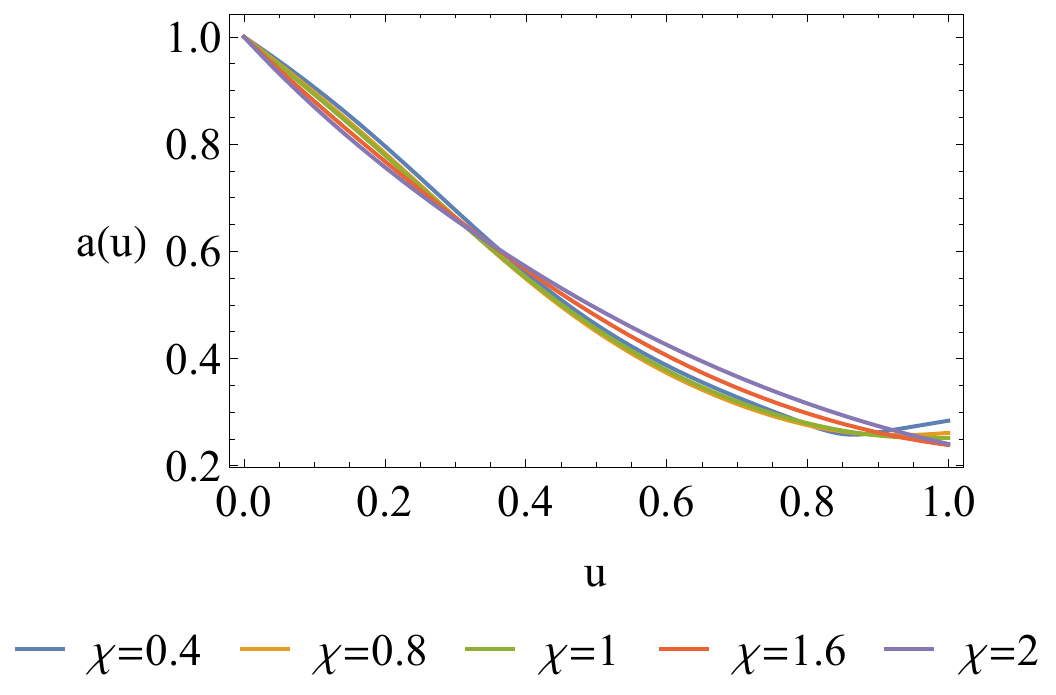}\ \hspace{0.8cm}
\includegraphics[scale=0.6]{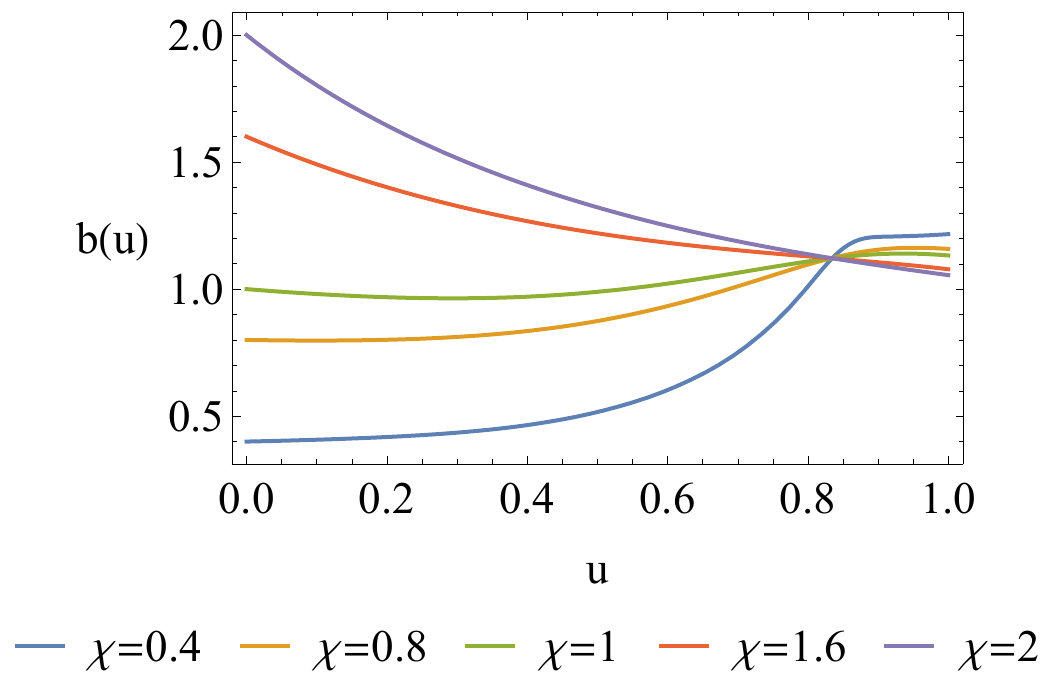}\ \\
\includegraphics[scale=0.6]{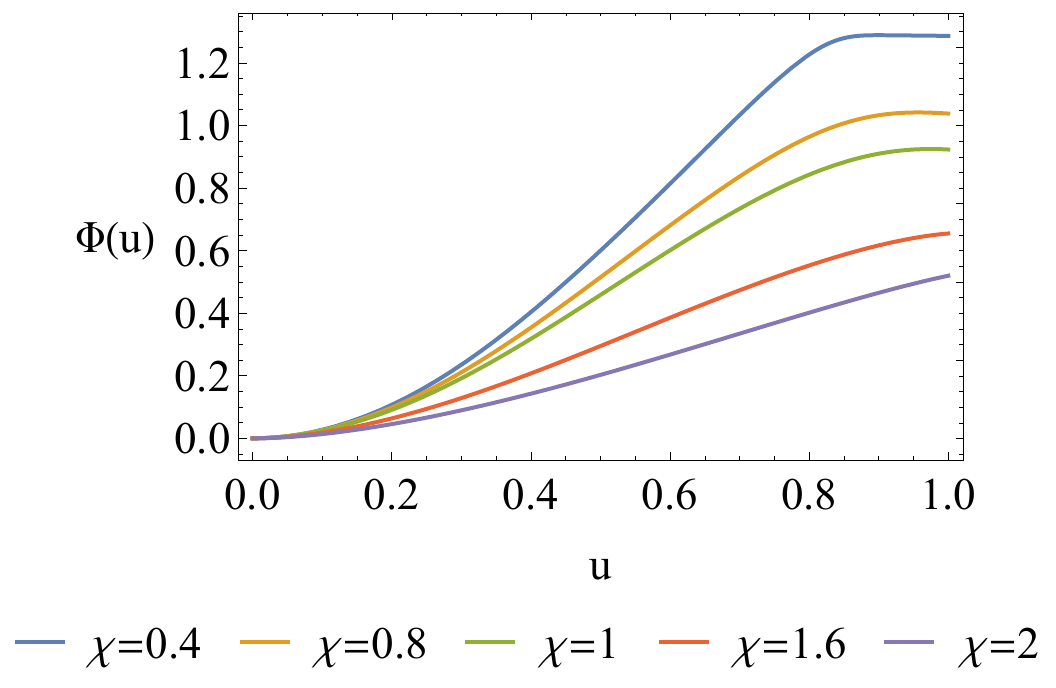}\ \\
\caption{\label{fig_phi} Plots of $U(u)$, $V(u)$, $a(u)$, $b(u)$ and $\Phi(u)$ as the function of $u$ with different doping $\chi$ at $T=0.01$.}}
\end{figure}

To numerically solve the EOMs, we need impose the boundary conditions at the UV boundary and the horizon.
At the UV boundary, $u\rightarrow 0$, we require that the geometry approaches $AdS_4$ with deformations corresponding to chemical potential $\mu$
and $\phi$ follows the behavior of Eq. \eqref{phiu-asy}.
At the horizon, $u\rightarrow 1$, the regular boundary conditions should be imposed.
FIG.\ref{fig_phi} shows the plots of $U(u)$, $V(u)$, $a(u)$, $b(u)$ and $\Phi(u)$ as the function of $u$ with different doping $\chi$ at $T=0.01$.
Some brief comments, which is helpful to understand the picture of the fermionic spectrum studied in following section, are summarized as what follows.
\begin{itemize}
  \item With the increase of $\chi$, the value of $\Phi(u)$ decreases at the horizon but vanishes near the UV boundary.
  It is a crucial ingredient in understanding the effect of the pseudoscalar coupling because it provides a direct coupling
  between scalar field and Dirac field.
  \item The value of the metric $V(u)$ increases at the horizon as $\chi$ increases.
  \item When we tune $\chi$, the function $b(u)$ greatly changes whether at the horizon or near the UV boundary.
  Especially, the change is very evident near the UV boundary.
\end{itemize}

\subsection{Dirac equation}

To study the fermionic spectrum, we consider the following Dirac action with Yukawa coupling between the spinor field
and the scalar field over this gravitational background \cite{Wu:2019orq}
\begin{subequations}
\begin{align}
S_{D}&=i\int d^{4}x \sqrt{-g}\overline{\zeta}\left(\Gamma^{a}\mathcal{D}_{a} - m_{\zeta} \right)\zeta\,,
\label{fermAction}
\
\\
S_{Y}&=\int d^{4}x \sqrt{-g}\left[
\eta_2\bar{\zeta}\Phi\Gamma^5\zeta+h.c.\right]\,.
\label{Yukawa}
\end{align}
\end{subequations}
$\Gamma^a=(e_{\mu})^a\Gamma^{\mu}$ and $(e_{\mu})^{a}$ are a set of orthogonal normal vector bases.
$\mathcal{D}_{a}=\partial_{a}+\frac{1}{4}(\omega_{\mu\nu})_{a}\Gamma^{\mu\nu}-iq A_{a}$
and $(\omega_{\mu\nu})_{a}$ is the spin connection 1-forms.
$\Gamma^5$ is the chirality matrix satisfying $\{\Gamma^5,\Gamma^{\mu}\}=0$.

From the above actions (\ref{fermAction}) and (\ref{Yukawa}), we have the following Dirac equation
\begin{eqnarray}
\label{DiracEquation1}
\Gamma^{a}\mathcal{D}_{a}\zeta-m_{\zeta}\zeta
-i\eta_2\Phi\Gamma^5\zeta=0.
\end{eqnarray}
In order to numerically solve the Dirac equation and read off the fermionic spectrum,
we make a redefinition of $\zeta=(g_{tt}g_{xx}g_{yy})^{-\frac{1}{4}}\mathcal{F}$
and expand $\mathcal{F}$ in momentum space as
\begin{eqnarray}
\mathcal{F}=\int\frac{d\omega dk}{2\pi}F(u,k)e^{-i\omega t + ikx},
\end{eqnarray}
where we have set $k_x=k$ and $k_y=0$.
In addition, we split the spinor into $F=(F_{1},F_{2})^{T}$ with $
F_{\alpha} \equiv (\mathcal{A}_{\alpha}, \mathcal{B}_{\alpha})^{T}
$
, $\alpha=1,2$,
and choose a specific gamma matrices as
\begin{eqnarray}
\label{GammaMatrices}
 && \Gamma^{3} = \left( \begin{array}{cc}
-\sigma^3  & 0  \\
0 & -\sigma^3
\end{array} \right), \;\;
 \Gamma^{0} = \left( \begin{array}{cc}
 i \sigma^1  & 0  \\
0 & i \sigma^1
\end{array} \right), \;\;
\Gamma^{1} = \left( \begin{array}{cc}
-\sigma^2  & 0  \\
0 & \sigma^2
\end{array} \right),  \cr
&&
 \Gamma^{2} = \left( \begin{array}{cc}
 0  & \sigma^2  \\
\sigma^2 & 0
\end{array} \right), \;\;
 \Gamma^{5} = \left( \begin{array}{cc}
 0  & i\sigma^2  \\
-i\sigma^2 & 0
\end{array} \right).
\end{eqnarray}
And then, we have the following Dirac equation of 4-component spinor
\begin{eqnarray}
\label{DiracEAB1}
\left(\frac{1}{\sqrt{g_{uu}}}\partial_{u}\mp m_{\zeta} \right)\left( \begin{matrix} \mathcal{A}_{1} \cr  \mathcal{B}_{1} \end{matrix}\right)
\pm (\omega+ q A_{t})\frac{1}{\sqrt{g_{tt}}}\left( \begin{matrix} \mathcal{B}_{1} \cr  \mathcal{A}_{1} \end{matrix}\right)
-\frac{k}{\sqrt{g_{xx}}} \left( \begin{matrix} \mathcal{B}_{1} \cr  \mathcal{A}_{1} \end{matrix}\right)
+ i\eta_2\Phi\left( \begin{matrix} \mathcal{B}_{2} \cr  \mathcal{A}_{2} \end{matrix}\right)
=0,
\nonumber
\\
\end{eqnarray}
\begin{eqnarray} \label{DiracEAB2}
\left(\frac{1}{\sqrt{g_{uu}}}\partial_{u}\mp m_{\zeta}\right)\left( \begin{matrix} \mathcal{A}_{2}\cr  \mathcal{B}_{2}\end{matrix}\right)
\pm (\omega+ q A_{t})\frac{1}{\sqrt{g_{tt}}}\left( \begin{matrix} \mathcal{B}_{2}\cr  \mathcal{A}_{2}\end{matrix}\right)
+\frac{k}{\sqrt{g_{xx}}} \left( \begin{matrix} \mathcal{B}_{2} \cr  \mathcal{A}_{2} \end{matrix}\right)
- i\eta_2\Phi\left( \begin{matrix} \mathcal{B}_{1} \cr  \mathcal{A}_{1} \end{matrix}\right)
=0.
\nonumber
\\
\end{eqnarray}

At the horizon, we need impose the ingoing boundary condition, which is
\cite{Wu:2019orq}
\begin{equation}
\left( \begin{matrix} \mathcal{A}_{\alpha}(u,\textbf{k}) \cr  \mathcal{B}_{\alpha}(u,\textbf{k}) \end{matrix}\right)
=c_\alpha\left( \begin{matrix} 1
\cr  -i\end{matrix}\right)(1-u)^{-\frac{i\omega}{4\pi T}}.
\end{equation}
Near the AdS boundary, the behavior of the Dirac field follows
\begin{eqnarray} \label{BoundaryBehaviour}
\left( \begin{matrix} \mathcal{A}_{\alpha} \cr  \mathcal{B}_{\alpha}\end{matrix}\right)
{\approx} a_{\alpha}u^{m_{\zeta}}\left( \begin{matrix} 1 \cr  0 \end{matrix}\right)
+b_{\alpha}u^{-m_{\zeta}}\left( \begin{matrix} 0 \cr 1 \end{matrix}\right).
\end{eqnarray}
By holography, we can read off the retarded Green function
\begin{eqnarray}
a_{\alpha}=G_{\alpha\alpha'}b_{\alpha'}.
\end{eqnarray}
Next we shall study the measurable spectral function $A(\omega,k_x,k_y)\sim Im(TrG)$.

\section{Numerical results}\label{sec-NR}

\begin{figure}
\center{
\includegraphics[scale=0.5]{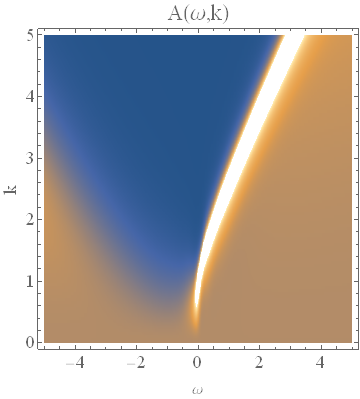}\ \hspace{0.8cm}
\includegraphics[scale=0.5]{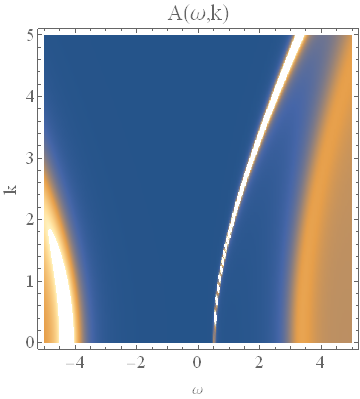}\ \\
\caption{\label{fig_dp} Density plots of spectral function $A(\omega,k)$ for $\chi=1$ at $T=0.01$.
Left plot is for $\eta_2=0$ and right plot for $\eta_2=4$.}}
\end{figure}
\begin{figure}
\center{
\includegraphics[scale=0.8]{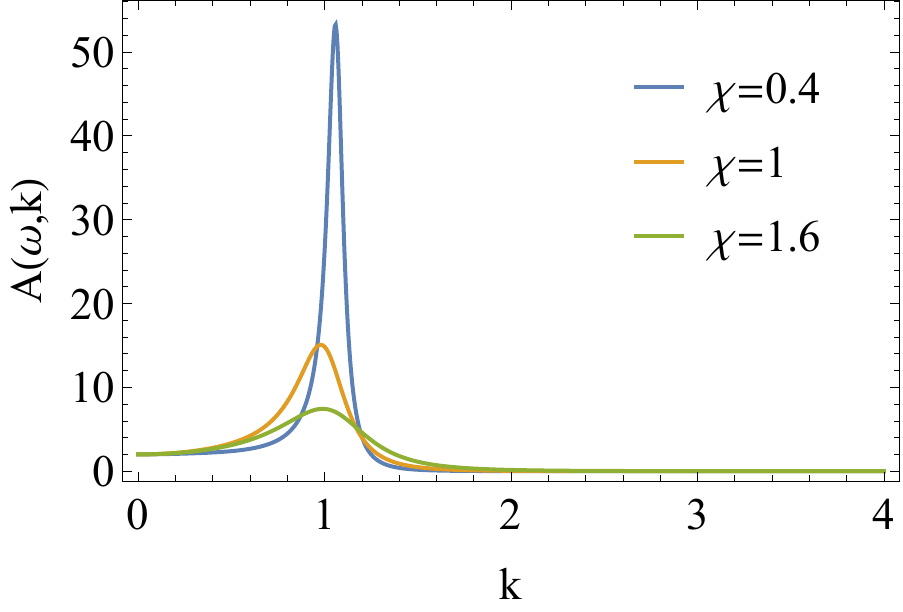}\ \\
\caption{\label{fig_Avsk} Spectral function $A(\omega,k)$ for $\eta_2=0$ and different $\chi$ at $T=0.01$.
}}
\end{figure}

In this section, we numerically study the holographic fermionic spectrum.
We only focus on the case of $m_{\zeta}=0$ and in most of case, we take $q=1$ except we make a special mark.
FIG.\ref{fig_dp} show the density plots of spectral function $A(\omega,k)$ for $\chi=1$ at $T=0.01$ (left panel is for $\eta_2=0$ and right panel for $\eta_2=4$).
When $\eta_2=0$, a small bump emerges near the Fermi level ($\omega=0$).
It is different from that over RN-AdS background for which a sharp peak shows up near $\omega=0$ \cite{Liu:2009dm}.
When we turn down the doping parameter $\chi$, the bump becomes a small peak (FIG.\ref{fig_Avsk}).
Conversely, when we turn up $\chi$, the bump becomes smaller (FIG.\ref{fig_Avsk}).
It indicates that the doping suppresses the bump at Fermi level.

\begin{figure}
\center{
\includegraphics[scale=0.8]{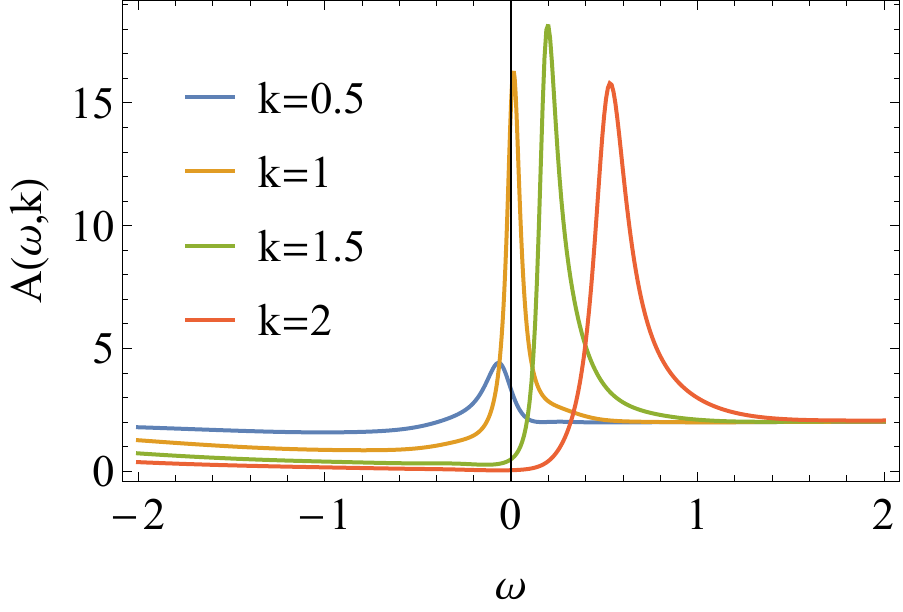}\ \hspace{0.8cm}
\includegraphics[scale=0.8]{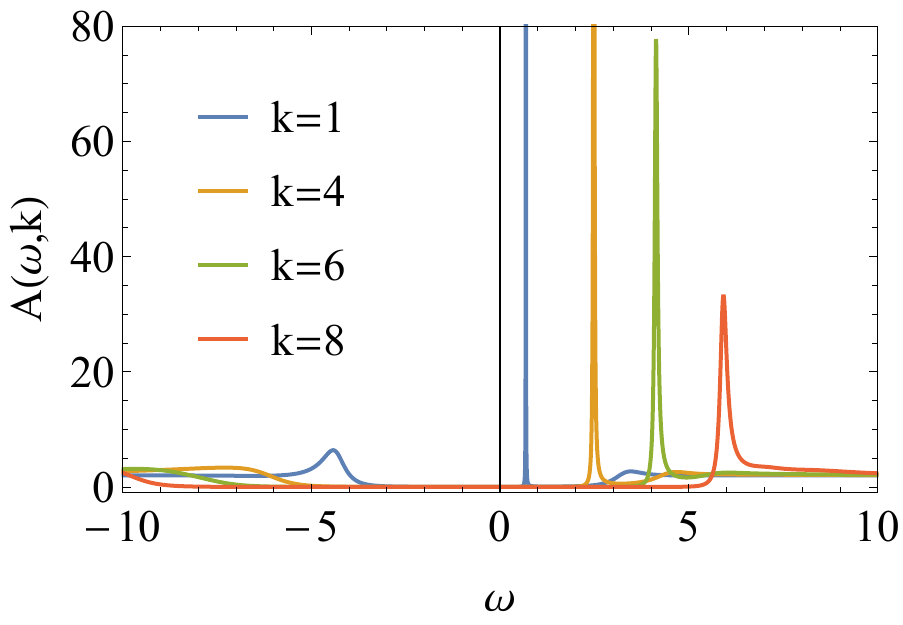}
\ \\
\caption{\label{fig_Avsomega} Spectral function $A(\omega,k)$ with $\chi=1$ at $T=0.01$ for sample $k$.
Left plot is for $\eta_2=0$ and right plot for $\eta_2=4$.
}}
\end{figure}

When we turn on the pseudoscalar Yukawa coupling $\eta_2$ and tune it large,
a gap emerges around $\omega=0$ (right panel in FIG.\ref{fig_dp}).
FIG.\ref{fig_Avsomega} further shows this process.
From FIG.\ref{fig_Avsomega}, we clearly see that for $\eta_2=0$, the peaks (bumps) distribute near $\omega=0$.
While for $\eta_2=4$, the peaks (bumps) disappears around $\omega=0$ and a gap emerges instead.
In particular, with the increase of $k$, the peaks (bumps) are pushed away from $\omega=0$, which confirms that the gap exists for all $k$.
\begin{figure}
\center{
\includegraphics[scale=0.8]{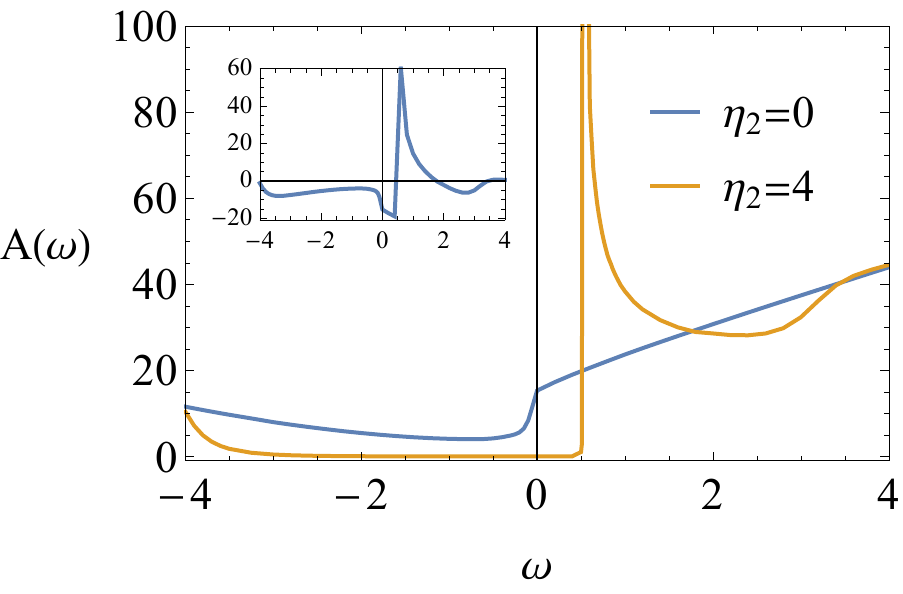}
\ \\
\caption{\label{fig_Avsomega_ik} DOS $A(\omega)$ as the function of $\omega$ for different $\eta_2$. Here $\chi=1$ and $T=0.01$.
The inset shows $A(\omega,\eta_2=4)-A(\omega,\eta_2=0)$ as a function of $\omega$.
}}
\end{figure}

The density of state (DOS) is an effective measure to study the properties of the gap.
It is defined as the integral of the spectral function $A(\omega,k)$ over $k$ space
and measures the total weight of the spectral function.
FIG.\ref{fig_Avsomega_ik} shows DOS $A(\omega)$ as the function of $\omega$ for different $\eta_2$.
We summarize the properties as what follows.
\begin{itemize}
  \item
  The spectrum rearranges at low frequency region (the absolute value $|\omega|$ is small),
  which is induced by the pseudoscalar Yukawa coupling. It attributes to the non-trivial profile of scalar field near horizon.
  \item As $\eta_2$ increases, there is the transfer of the spectral weight from the low energy band ($\omega<0$) to the high energy band ($\omega>0$).
  The transfer is over low energy scales but not over all energy scales.
  It is different from the Mott physics and also that in RN-AdS background, for which the spectral weight transfers over all energy scales \cite{Edalati:2010ww,Edalati:2010ge}.
  From FIG.\ref{fig_Avsomega_ik}, especially the inset of FIG.\ref{fig_Avsomega_ik} which exhibits the difference between $A(\omega,\eta_2=4)$ and $A(\omega,\eta_2=0)$,
  we see that when $|\omega|$ is large (the high frequency region),
  the DOS for all $\eta_2$ approaches the same value.
  It is because the scalar field vanishes on the UV boundary.
\end{itemize}
These properties are the same as that revealed in \cite{Wu:2019orq}.
Therefore, we conclude that even when the doping is introduced in this system,
it do not change the basic properties from pseudoscalar Yukawa coupling.

\begin{figure}
\center{
\includegraphics[scale=0.6]{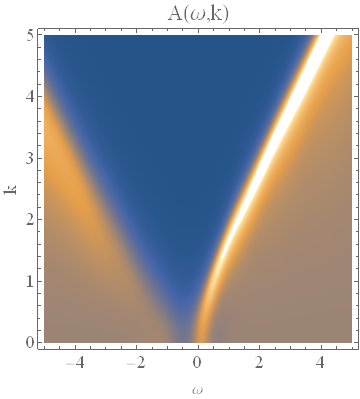}
\ \\
\caption{\label{fig_DP_tem} The density plot of spectral function $A(\omega,k)$ for $\chi=1$ and $\eta_2=4$ at $T=0.15$.
}}
\end{figure}
When we heat this fermionic system with doping, we find that the gap closes (see FIG.\ref{fig_DP_tem}).
It is a universal property shared by Mott physics, the fermionic system with dipole coupling \cite{Edalati:2010ww,Edalati:2010ge},
and also that from pseudoscalar Yukawa coupling \cite{Wu:2019orq}.

\begin{figure}
\center{
\includegraphics[scale=0.55]{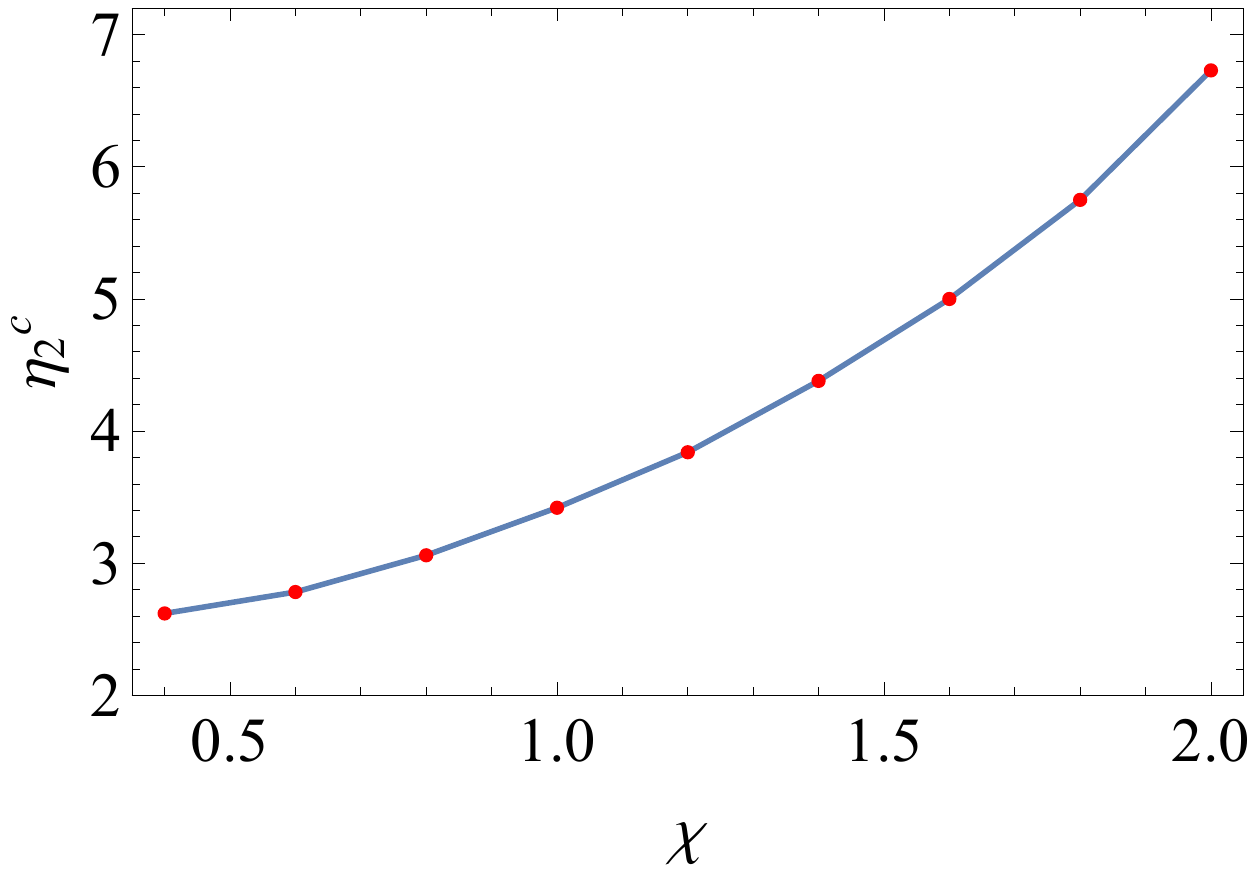}\ \hspace{0.8cm}
\includegraphics[scale=0.58]{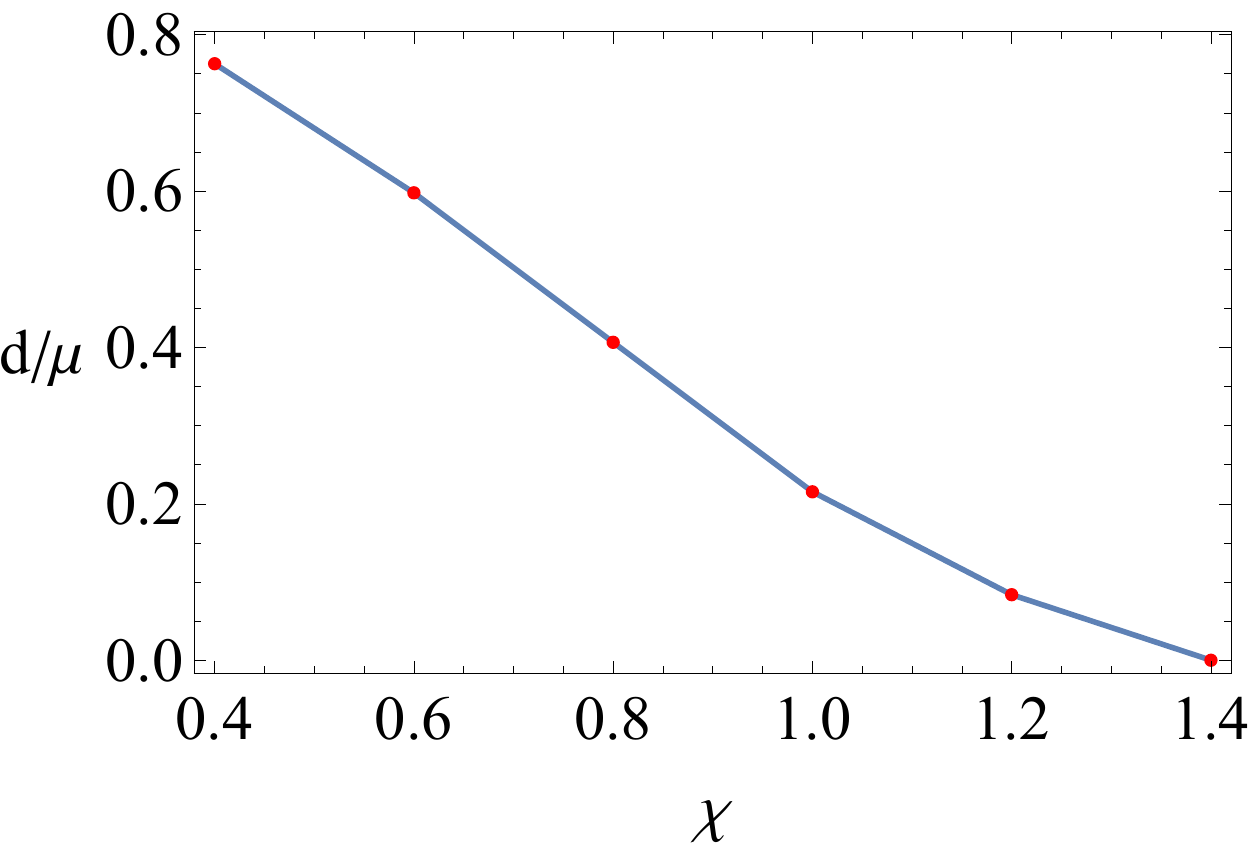}\ \\
\caption{\label{fig_etac} Left plot: The relation between the doping $\chi$ and $\eta_2^c$ at $T=0.01$.
Right plot: The relation between the doping $\chi$ and the width of the gap $d/\mu$ for fixed $\eta_2=4$ at $T=0.01$.
}}
\end{figure}

Now, we turn to quantitatively study the effect from the doping.
To this end, we introduce the critical value of the onset of the gap $\eta_2^c$ to signal the formation of the gap.
It is identified as that the DOS at the Fermi level drops below some small number.
In numerical calculation, we take this number as $10^{-3}$.
Left plot in FIG.\ref{fig_etac} and TABLE \ref{table_etac} show the relation between the doping $\chi$ and $\eta_2^c$.
The study indicates that when the doping becomes large, the gap emerges more difficult.
Note that with the increase of $\chi$, the amplitude of the scalar field near the horizon become more evident (see FIG.\ref{fig_phi} and the associated comments in section \ref{subsec-bg}),
which usually make the gap open more easily as observed in \cite{Wu:2019orq} because the low frequency spectrum probes the near horizon geometry.
But here we observe a contrary result. It indicates that there is a competition between the pseudoscalar Yukawa coupling and the doping.
The essential reason may be from the background geometry and in particular the profiles of the gauge field $B$, which are greatly changed by the doping (see FIG.\ref{fig_phi}).
But deeper understanding on this issue, we need to implement an analytical study.
We leave it for future work.

\begin{widetext}
\begin{table}[ht]
\begin{center}
\begin{tabular}{|c|c|c|c|c|c|c|c|}
         \hline
~$\chi$~ &~$0.4$~&~$0.8$~&~$1$~&~$1.6$~&~$2$~
          \\
        \hline
~$\eta_{2}^c$~ & ~$2.620$~&~$3.06$~&~$3.420$~&~$5.000$~&~$6.730$~
          \\
        \hline
\end{tabular}
\caption{\label{table_etac} $\eta_2^c$ with different $\chi$ at $T=0.01$.
}
\end{center}
\end{table}
\end{widetext}

Also we study the relation between the doping $\chi$ and the width of the gap $d/\mu$ for fixed $\eta_2=4$ at $T=0.01$ (right plot in FIG.\ref{fig_etac}).
We find that with the increase of the doping $\chi$, the size of the gap becomes smaller.
When $\chi\geq 1.4$, the gap vanishes.
To understand the physics behind this phenomenon, it need call for an analytical treatment, for which we leave for future work.
Here, we only provide a brief comment.
The size of gap is mainly determined by the near horizon geometry.
From the right plot in FIG.\ref{fig_etac}, we see that besides the amplitude of the scalar field near the horizon, the amplitudes of the other fields,
including the gauge fields, $A$ and $B$, and the metric field $V$, have great variations as the doping $\chi$ increases.
Therefor, there are several elements impacting on the size of the gap.

\begin{figure}
\center{
\includegraphics[scale=0.65]{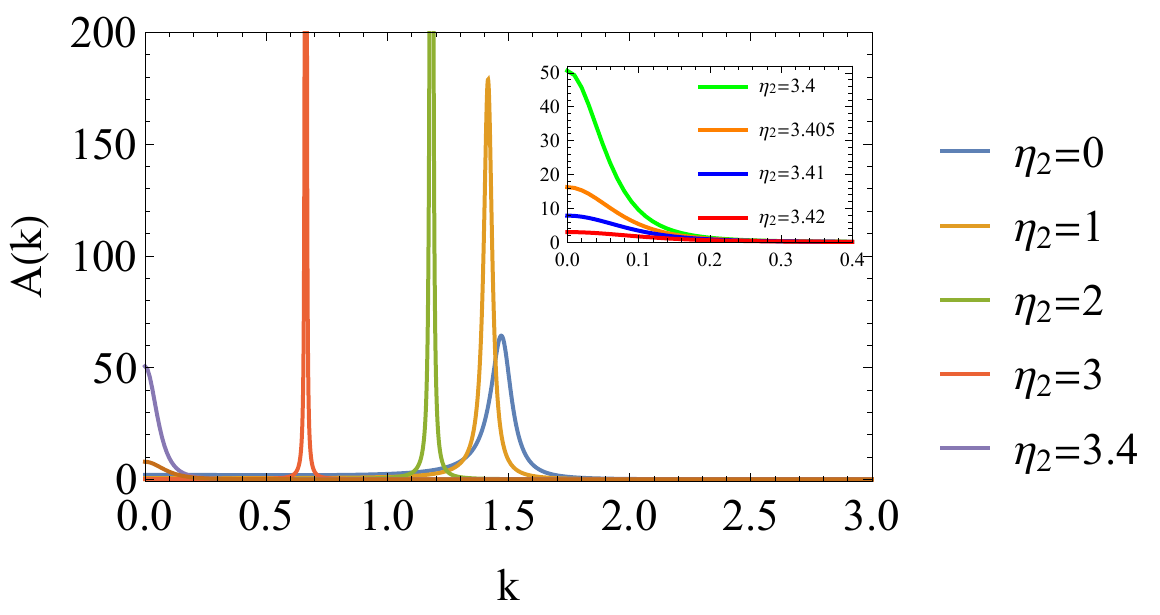}\ \hspace{0.1cm}
\includegraphics[scale=0.65]{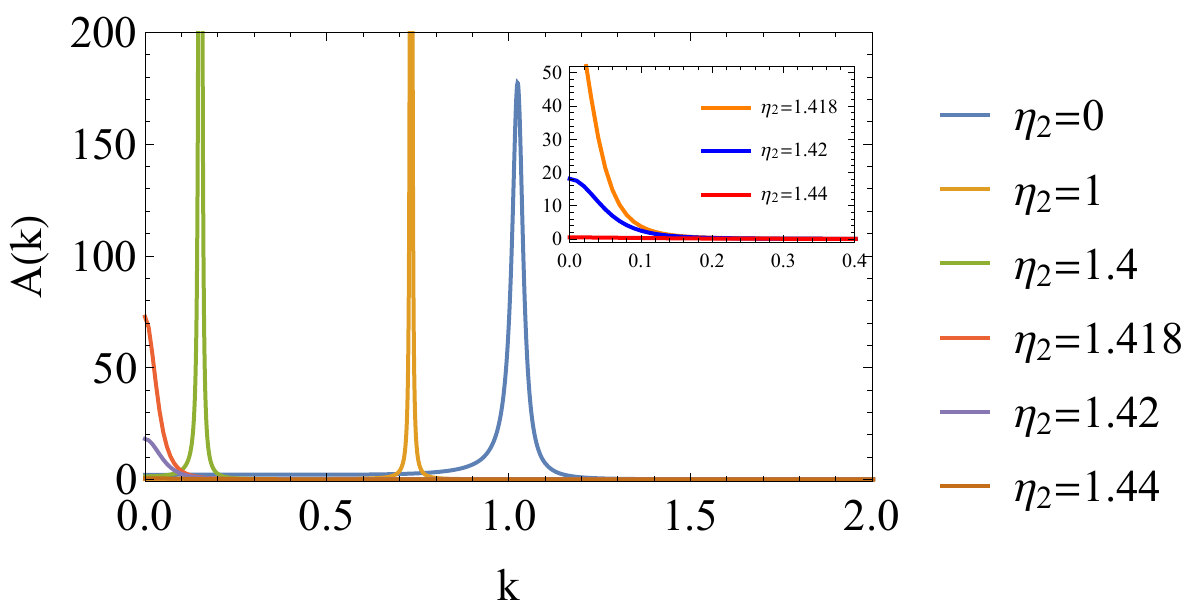}
\ \\
\caption{\label{fig_Avskq1p2} Left plot: Fermionic spectrum $A(k)$ as the function of $k$ for $\chi=1$, $q=1.2$ and $T=0.01$.
Right plot: Fermionic spectrum $A(k)$ as the function of $k$ over the background with one gauge field studied in Ref.\cite{Wu:2019orq}.
Here we take $q=1.2$ and $T=0.01$ and other parameters are the same as in \cite{Wu:2019orq}.
}}
\end{figure}
\begin{figure}
\center{
\includegraphics[scale=0.8]{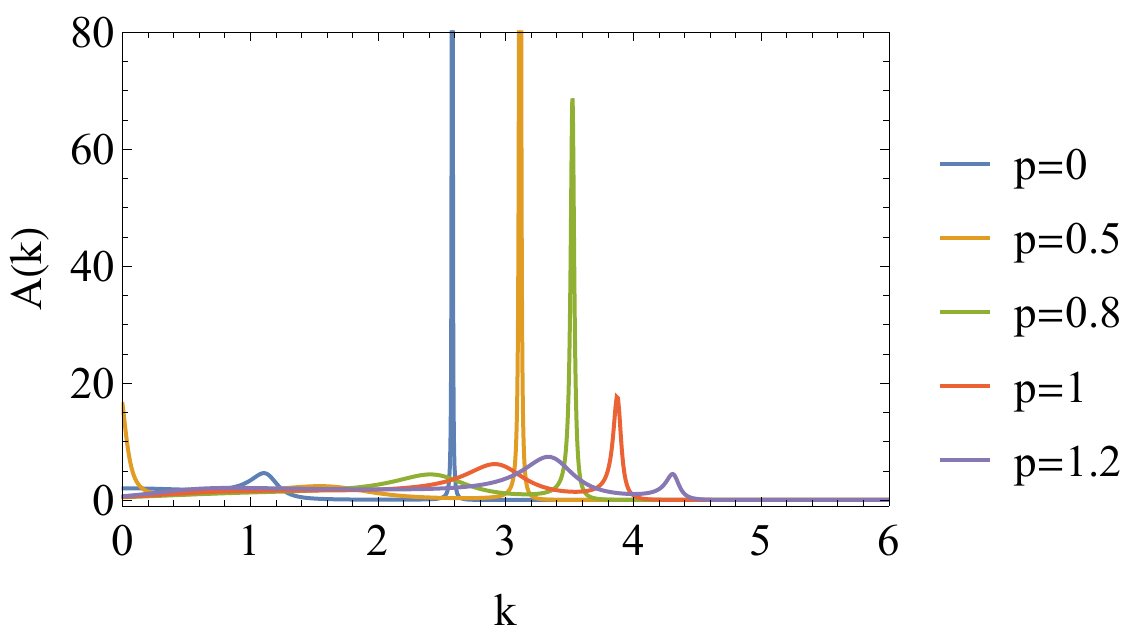}
\ \\
\caption{\label{fig_Avskq1p2dipole} Fermionic spectrum $A(k)$ with dipole coupling proposed in \cite{Edalati:2010ww,Edalati:2010ge}
as the function of $k$ over RN-AdS black brane. Here we take $q=1.2$ and $T=0.01$.
}}
\end{figure}

Before closing this section, we would like to present a brief discussion on how the gap emerges.
As pointed out in the introduction, the formation of the gap is not from the localization but only from the result of increasing Yukawa coupling.
The left plot in FIG.\ref{fig_Avskq1p2} shows the spectrum $A(\omega)$ as the function of $k$ at the Fermi level $\omega=0$.
From this figure, we see that when $\eta_2=0$, a sharp Fermi peak emerges.
As the Yukawa coupling increases, this peak becomes sharper and closer the position of $\omega=0$.
However, as the Yukawa coupling further increases and is beyond certain value, the peak hits the vertical axis and becomes a small bump around $\omega=0$.
After that, as the Yukawa coupling further increases, the bump shrinks to zero at $\omega=0$ and the gap comes into being.
There is the same property for the Yukawa coupling in the background with one gauge field studied in \cite{Wu:2019orq} (see the right plot in FIG.\ref{fig_Avskq1p2}).
However, for the bulk dipole coupling proposed in \cite{Edalati:2010ww,Edalati:2010ge}, we find that as the dipole coupling increases,
the peaks shrinks to zero and is pushed off the position of $\omega=0$ (see FIG.\ref{fig_Avskq1p2dipole}), which results in the formation of the gap.

\section{Conclusion and discussion}\label{sec-con}

In this paper, we introduce a two-current model, which includes two gauge fields and a neutral scalar field.
This model supports an AdS black brane geometry with scalar hair.
Two tuneable chemical potentials are introduced,
which induce unbalance of numbers and so introduce a controllable doping parameter.
The study on the fermionic response shows that with the increase of the doping,
the gap opens more difficult. This result indicates that there is a competition between the pseudoscalar Yukawa coupling and the doping.
It attributes to greatly change of the background geometry and the profiles of the gauge fields as well as scalar field causing by the doping.

We would like to point out that the features of the electronic spectrum in our present model studied here are not related to the features of electric transport.
To consider the contribution to the electric transport from probe fermionic system in the gravity background,
we shall perform a one-loop calculation on the gravity side as Refs. \cite{Faulkner:2010zz,Faulkner:2013bna},
in which the authors find that a consistent picture from the electric transport and the fermionic spectrum.
In particular, the transport lifetime is consistent with the single-particle lifetime.
In addition, in the case where the fermionic spectrum exhibits a marginal Fermi liquid behavior, the resistivity varies linearly with temperature,
which are both the important characteristics of strange metal.
Along this direction, when the pseudoscalar Yukawa coupling is introduced,
it is interesting to explore the electric transport by performing a one-loop calculation on the gravity side and to see if we have a consistent picture
from the electric transport and the fermionic spectrum.

\begin{acknowledgments}

We would like to thank anonymous referee for his/her nice suggestions and comments.
This work is supported by the Natural Science Foundation of China under
Grant Nos. 11775036, 11905182 and Fok Ying Tung Education Foundation under Grant No. 171006.
J. P. Wu is also supported by Top Talent Support Program from Yangzhou University.

\end{acknowledgments}

\end{document}